\documentclass[conference,10pt]{IEEEtran}
\usepackage[cmex10]{amsmath}
\usepackage{amsfonts}
\usepackage[pdftex]{color,graphicx}
\usepackage{subfig}
\usepackage{latexsym}

\usepackage{xspace}
\newtheorem{thm}{Theorem}[section]  
\newtheorem{corol}[thm]{Corollary}
\newtheorem{lem}[thm]{Lemma}
\newtheorem{prop}[thm]{Proposition}
\newtheorem{defn}{Definition}[section]  
\newtheorem{eg}{Example}[section]  

\newcounter{MYtempeqncounter}
\newcounter{MYtempeqncounter1}
\newcounter{MYtempeqncounter2}

\newcommand{\url}[1]{{#1}}

\newcommand{\namedref}[2]{#1~\ref{#2}}
\newcommand{\sectionref}[1]{\namedref{Section}{sec:#1}}
\newcommand{\theoremref}[1]{\namedref{Theorem}{thm:#1}}

\newcommand{\corollaryref}[1]{\namedref{Corollary}{cor:#1}}

\newcommand{\lemmaref}[1]{\namedref{Lemma}{lem:#1}}
\newcommand{\propositionref}[1]{\namedref{Proposition}{prop:#1}}

\newcommand{\figureref}[1]{\namedref{Figure}{fig:#1}}

\newcommand{\appendixref}[1]{\namedref{Appendix}{app:#1}}

\newcommand{\TODO}[1]{\textcolor{red}{#1}}

\newcommand{\defineqq}{\stackrel{\triangle}{=}}
\newcommand{\namedeqq}[1]{\ensuremath{\stackrel{\mathrm (#1)}{=}}}

\newcommand{\Real}{\ensuremath{\mathbb{R}}\xspace}
\newcommand{\Rplus}{\ensuremath{{{\mathbb{R}}_+}}\xspace}

\newcommand{\Com}[1]{\ensuremath{{\mathrm C}_\text{\sf #1}}}
\newcommand{\GWlower}{\ensuremath{\mathcal L}_\text{\sf GW}}

\newcommand{\ihull}[1]{\ensuremath{i\left(#1\right)}}

\newcommand{\A}{\ensuremath{\text{\sf A}}}
\newcommand{\B}{\ensuremath{\text{\sf B}}}
\newcommand{\C}{\ensuremath{\text{\sf C}}}
\newcommand{\CI}{\ensuremath{\text{\sf CI}}}
\newcommand{\RD}{\ensuremath{\text{\sf RD}}}

\newcommand{\RDO}{\ensuremath{\text{\sf RD-0}}}
\newcommand{\sX}{\ensuremath{\mathcal X}}
\newcommand{\sY}{\ensuremath{\mathcal Y}}

\newcommand{\sU}{\ensuremath{\mathcal U}}

\newcommand{\bbZ}{\ensuremath{\mathbb Z}}

\newcommand{\sRsRD}{\ensuremath{{\mathcal R}_\RD}}

\newcommand{\Wyner}{\ensuremath{\sf Wyner}}
\newcommand{\ACI}{\ensuremath{\text{\sf ACI}}}
\newcommand{\GW}{\ensuremath{\text{\sf GW}}\xspace}
\newcommand{\GKW}{\ensuremath{\text{\sf GK}}\xspace}

\newcommand{\sRsGW}{\ensuremath{{\mathcal R}_\GW}}
\newcommand{\sRsGKW}{\ensuremath{{\mathcal R}_\ACI}}
\newcommand{\sP}{\ensuremath{{\mathcal P}}}
\newcommand{\bU}{\ensuremath{{\mathbf U}}}

\newcommand{\Kfunc}{\ensuremath{\sRsGKW}\xspace}
\newcommand{\KK}[2]{\ensuremath{\Kfunc({#1},{#2})}\xspace}

\newcommand{\tp}{\ensuremath{\tilde{p}}}

\newcommand{\Pialice}{\ensuremath{\Pi^{\mathrm{view}}_{\mathrm{Alice}}}\xspace}
\newcommand{\Pibob}{\ensuremath{\Pi^{\mathrm{view}}_{\mathrm{Bob}}}\xspace}
\newcommand{\Pialiceout}{\ensuremath{\Pi^{\mathrm{out}}_{\mathrm{Alice}}}\xspace}
\newcommand{\Pibobout}{\ensuremath{\Pi^{\mathrm{out}}_{\mathrm{Bob}}}\xspace}

\begin{document}

\title{Assisted Common Information: Further Results}

\author{\IEEEauthorblockN{Vinod M. Prabhakaran}\\\IEEEauthorblockA{\'Ecole
Polytechnique F\'ed\'erale de Lausanne\\ Switzerland} \and \IEEEauthorblockN{Manoj M. Prabhakaran}\\
\IEEEauthorblockA{University of Illinois, Urbana-Champaign\\
Urbana, IL 61801}}


\maketitle

\begin{abstract}
{We presented assisted common information as a generalization of
G\'{a}cs-K\"{o}rner (GK) common information at ISIT 2010.  The motivation
for our formulation was to improve upperbounds on the efficiency of
protocols for secure two-party sampling (which is a form of secure
multi-party computation). Our upperbound was based on a monotonicity
property of a rate-region (called the assisted residual information region)
associated with the assisted common information formulation.

In this note we present further results. We explore the connection of
assisted common information with the Gray-Wyner system. We show that the
assisted residual information region and the Gray-Wyner region are
connected by a simple relationship: the assisted residual information
region is the increasing hull of the Gray-Wyner region under an affine map.
Several known relationships between GK common information and Gray-Wyner
system fall out as consequences of this. Quantities which arise in other
source coding contexts acquire new interpretations.

In previous work we showed that assisted common information can be used to
derive upperbounds on the rate at which a pair of parties can {\em securely
sample} correlated random variables, given correlated random variables from
another distribution. Here we present an example where the bound derived
using assisted common information is much better than previously known
bounds, and in fact is tight. This example considers correlated random
variables defined in terms of standard variants of oblivious transfer, and
is interesting on its own as it answers a natural question about these
cryptographic primitives.
}

\end{abstract}

\section{Introduction}

If $U,V,W$ are independent random variables, a natural measure of ``common
information'' of $X=(U,V)$ and $Y=(U,W)$ is $H(U)$. Observers of either $X$
or $Y$ may produce the common part $U$ and conditioned on this common part,
there is no residual information, i.e., $I(X;Y|U)=0$.
G\'acs-K\"orner (GK) common
information~\cite{GacsKo73,Witsenhausen75} is a generalization of this to
arbitrary $X,Y$. Two observers see $X^n=(X_1,X_2,\ldots,X_n)$ and
$Y^n=(Y_1,Y_2,\ldots,Y_n)$, resp., where $(X_i,Y_i)$ are independent draws
of $(X,Y)$. The observers produce $W_1=f_1(X^n)$ and $W_2=f_2(X^n)$ which
have an asymptotically vanishing probability of not matching. GK common
information is the largest entropy rate (normalized by $n$) of such a
common random variable. It was however shown that this value is the largest
$H(U)$ for which the random variables can be written as $X=(U,V)$ and
$Y=(U,W)$ (where $U,V,W$ may be dependent), i.e., the definition captures
only an explicit form of common information in a single instance of $X,Y$.

At ISIT 2010 we presented a generalization of GK common
information~\cite{PrabhakaranPr10}. In our setup (see \figureref{common}),
an omniscient genie (who has access to the $X$ and $Y$ sequences) assists
the users in generating the common random variables by sending them
messages over rate-limited noiseless links. A three-dimensional trade-off
region which characterizes the trade-off between the rates of the two
noiseless links and the resulting residual information (defined as the
conditional mutual information between the source sequences conditioned on
the common random variable normalized by the length of the sequence) was
derived. We call this the {\em assisted residual information region}. When
the links have zero rates, we recover GK common information.

Our motivation for this generalization was an application to cryptography.
Distributed dependent random variables are an important resource in the
cryptographic task of secure multi-party computation. A fundamental problem
here is for two parties to securely generate a certain pair of random
variables, given another pair of random variables, by means of a protocol.
Our main result there was that the assisted residual dependency region of
the views of two parties engaged in such a protocol can only monotonically
expand and not shrink which immediately leads to upperbounds on the
efficiency with which a target pair of random variables can be generated
from another pair. This work generalized previous work on
monotones~\cite{WolfWu08}. These works are in the same vein as
\cite{Beaver96,DodisMi99,WinterNaIm03,ImaiMuNaWi04,ImaiMoNaWi06,ImaiMoNa06,CsiszarAh07,WinklerWu09}
which employ information theory to derive bounds on efficiency in
cryptography.

In the first part of this paper we explore connections between the assisted
common information system and the Gray-Wyner source coding system of
\cite{GrayWy74}. In the Gray-Wyner system, a pair of sources is decomposed
into three components: one public and two private. Using the public and one
of the private components, one of the pair of sources must be recoverable,
while the other source must be recoverable using the other private
component and the public component. Gray-Wyner region is a
three-dimensional region which characterizes the trade-offs between the
rates at which the three components can be encoded.

We show that the assisted residual information region and the Gray-Wyner
region are connected by a simple relationship: the assisted residual
information region is the increasing hull\footnote{Increasing hull $i(S)$
of a set $S\subseteq \Real^d$ is the set of all $s\in\Real^d$ such that
there is a $s'\in S$ such that $s\geq s'$, where the inequality is
component-wise.} of the Gray-Wyner region under an affine map.  Several
known relationships between GK common information and Gray-Wyner system
fall out as consequences of this. This also leads to alternative
interpretations (in terms of the assisted common information system) to
quantities which arise naturally in certain other source coding contexts.
However, it must be noted that the Gray-Wyner region itself does not
possess the monotonicity property which makes it less-suited for the
cryptographic application which motivated \cite{PrabhakaranPr10}.

The second half of the paper is a sequel to the cryptographic application
in \cite{PrabhakaranPr10}. There we showed an example where our upperbound
(on the efficiency with which a pair of random variables can be securely
generated from another pair) strictly improved upon bounds from previous
results. That example was contrived to highlight the shortcomings of prior
work. Here we give yet another example where the upperbound from our result
strictly improves on the prior work, but is further interesting for two
reasons: firstly, the new example is based on natural correlated random
variables that are widely studied (namely, variants of oblivious transfer),
and secondly the new upperbound we can prove actually matches an easy
lowerbound and is therefore tight. 

\section{Preliminaries}

\subsection{Assisted Common Information System}
We presented the following generalization of GK common
information at ISIT,~2010~\cite{PrabhakaranPr10}. 
We call it the assisted common information system.
\begin{figure}[tb]
\centering
\scalebox{0.23}{\includegraphics{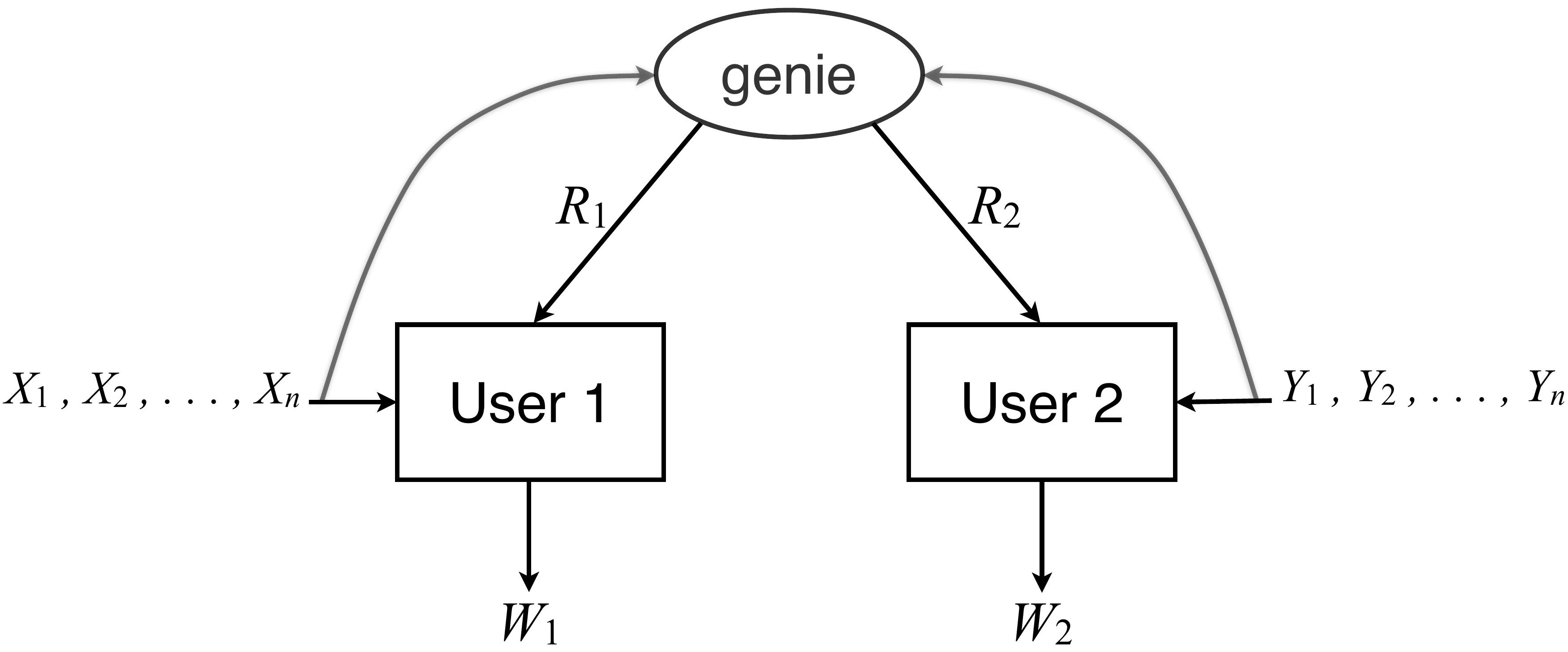}}\\
\caption{Setup for
assisted common information 
system. The users generate $W_1$ and $W_2$ which are required to agree with
high probability. A genie assists the users by sending separate messages to
them over rate-limited noiseless links. When the genie is absent the setup
reduces to the one for G\'{a}cs-K\"{o}rner common information.}
\label{fig:common}
\end{figure}
\begin{figure}[tb]
\centering
\scalebox{0.23}{\includegraphics{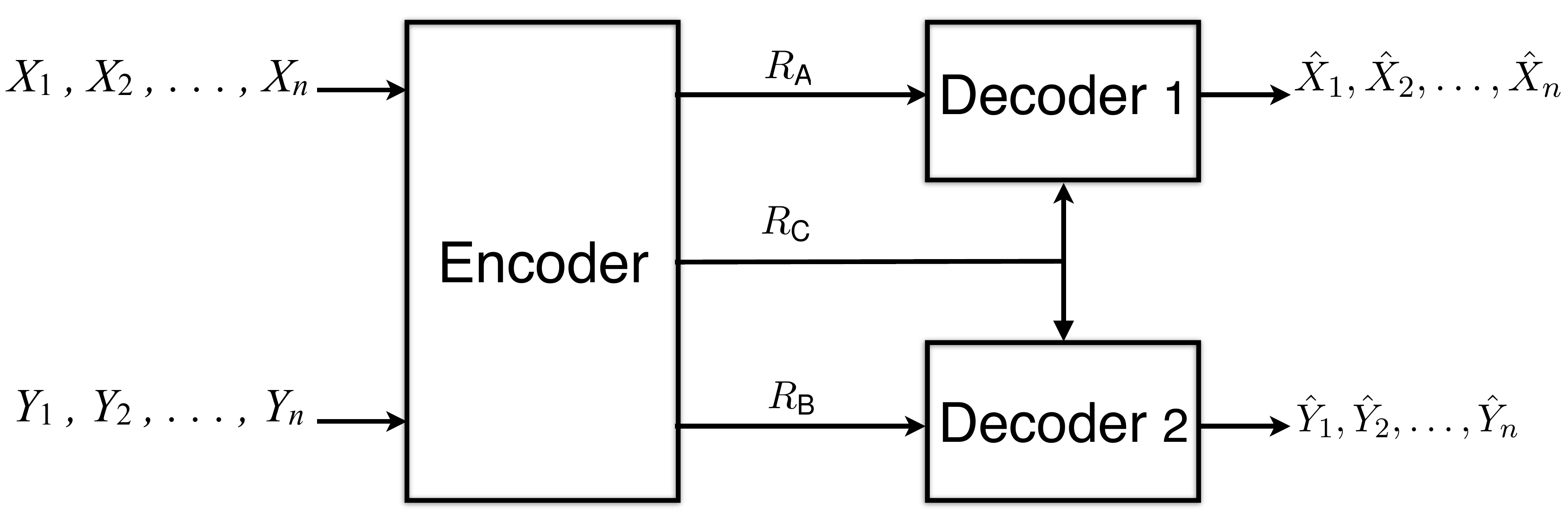}}\\
\caption{Setup for
Gray-Wyner (GW) system.}
\label{fig:GW}
\end{figure}

Consider \figureref{common}. For a pair of random variables $(X,Y)$, we say
that a rate pair $(R_1,R_2)$ {\em enables} a residual information rate
$R_\RD$ if for every $\epsilon>0$, there is a large enough integer $n$ and
(deterministic) functions $f_k:\sX^n\times\sY^n \rightarrow \{1,\ldots,
2^{n(R_k+\epsilon)}\}$, $(k=1,2)$, $g_1:\sX^n\times\{1,\ldots,
2^{n(R_1+\epsilon)}\} \rightarrow \bbZ$, and $g_2:\sY^n\times\{1,\ldots,
2^{n(R_2+\epsilon)}\} \rightarrow \bbZ$ (where $\bbZ$ is the set of
integers) such that
\begin{align}
&\Pr\left( g_1(X^n,f_1(X^n,Y^n)) \neq g_2(Y^n,f_2(X^n,Y^n)) \right) \leq
\epsilon,\label{eq:prob-of-error}\\
&\frac{1}{n} I(X^n;Y^n|g_1(X^n,f_1(X^n,Y^n))) \leq R_\RD + \epsilon.
\label{eq:RDrate}
\end{align}
\begin{defn} 
We define the {\em assisted residual information region}\footnote{We may
also define an analogous {\em assisted common information} region by
replacing the definition in \eqref{eq:RDrate} by \begin{align*}
\frac{1}{n}I(X^n,Y^n;g_1(X^n,f_1(X^n,Y^n))) \geq R_\CI - \epsilon.
\end{align*} See~\cite{PrabhakaranPr10} for this and its connection to the above
definition. In effect, the definitions are equivalent as we discuss there.
We work with assisted residual information region since it has a simple
monotonicity property (\theoremref{secure-realization-rate}) which makes it appealing for
deriving bounds for secure two-party sampling.} $\sRsGKW(X,Y)$ of a pair of
random variables $(X,Y)$ with joint distribution $p_{X,Y}$ as the set of
all $(r_1,r_2,r_\RD)\in \Rplus^3$ for which there is a $(R_1,R_2,R_\RD)$
such that $r_1\geq R_1$, $r_2\geq R_2$, $r_\RD\geq R_\RD$, and $(R_1,R_2)$
enables the residual information rate $R_\RD$. In other words,
\begin{align*}
\sRsGKW(X,Y)\defineqq
\ihull{\{(R_1,R_2,R_\RD): (R_1,R_2) \text{ enables } R_\RD\}},
\end{align*}
where $\ihull{S}$ denotes the {\em increasing hull} of $S\subseteq \Rplus^3$: 
$\ihull{S}=\{s\in\Rplus^3: s\geq s'\text{ component-wise for some }
 s'\in S \}$.
\end{defn} 
We will write $\sRsGKW$ when the random variables involved are obvious from
the context.

When the two rates from the genie are zero, we recover
G\'{a}cs-K\"{o}rner common information,
$\Com\GKW$~\cite{GacsKo73,Witsenhausen75}. Let $R_\RDO \defineqq \inf
\{R_\RD: (0,0,R_\RD)\in \sRsGKW(X,Y)\}$.  Then we have the following
proposition.
\begin{prop} \label{prop:GacsKo}
\begin{align}
\Com\GKW(X,Y) &= I(X;Y) - R_\RDO. \label{eq:CGKWfromRDO}\\
\intertext{Further}
R_\RDO &= \inf_{p_{U|XY}:I(X;U|Y)=I(Y;U|X)=0} I(X;Y|U)\label{eq:RDO}\\
\intertext{which gives}
\Com\GKW(X,Y) &= \sup_{p_{U|XY}:I(X;U|Y)=I(Y;U|X)=0} H(U). \label{eq:CGKW}
\end{align}
Moreover, $\Com\GKW(X,Y)=0$ unless there are $X',Y',U'$ such that $X=(X',U'),
Y=(Y',U')$, in which case $\Com\GKW=\max_{U':X=(X',U'),Y=(Y',U')} H(U')$.
\end{prop}

The proof of this proposition and all other results are available in the
appendix. The proof of \eqref{eq:RDO} relies on the following
characterization of $\sRsGKW$ which was proved in~\cite{PrabhakaranPr10}.
Let $\sP$ be the set of all marginal p.m.f's $p_{U|X,Y}$ such that the
cardinality of alphabet $\sU$ of $U$ is $|\sX||\sY|+2$.
\begin{prop}\label{prop:GKW}
\begin{align*}
&\sRsGKW(X,Y)=\\ 
&\;\ihull{\bigcup_{p_{U|X,Y}\in\sP}\{(I(Y;U|X),I(X;U|Y),I(X;Y|U))\}}
\end{align*}
\end{prop}

\subsection{Gray-Wyner system}

The Gray-Wyner system is shown in Figure~\ref{fig:GW}. It is a source coding
problem formulated as follows:  We say that a rate 3-tuple $(R_\A,R_\B,R_\C)$
is {\em achievable} if for every $\epsilon>0$, there is a large enough integer
$n$ and (deterministic) encoder functions $f_\A:\sX^n\times\sY^n \rightarrow
\{1,\ldots, 2^{n(R_\A+\epsilon)}\}$, $f_\B:\sX^n\times\sY^n \rightarrow
\{1,\ldots, 2^{n(R_\B+\epsilon)}\}$, $f_\C:\sX^n\times\sY^n \rightarrow
\{1,\ldots, 2^{n(R_\C+\epsilon)}\}$, and (deterministic) decoder functions
$g_{\A\C}:\{1,\ldots,2^{n(R_\A+\epsilon)}\} \times \{1,\ldots,
2^{n(R_\C+\epsilon)}\} \rightarrow \sX^n$, and
$g_{\B\C}:\{1,\ldots,2^{n(R_\B+\epsilon)}\} \times \{1,\ldots,
2^{n(R_\C+\epsilon)}\} \rightarrow \sY^n$ such that
\begin{align}
&\Pr\left( g_{\A\C}(f_\A(X^n,Y^n),f_\C(X^n,Y^n)) \neq X^n) \right) \leq
\epsilon,\\
&\Pr\left( g_{\B\C}(f_\B(X^n,Y^n),f_\C(X^n,Y^n)) \neq Y^n) \right) \leq
\epsilon.
\end{align}
\begin{defn} 
The Gray-Wyner region $\sRsGW(X,Y)$ is the set of all achievable rate 3-tuples.
\end{defn}
We write $\sRsGW$ when the random variables are clear from the context. 
A simple lower-bound to $\sRsGW(X,Y)$ is 
\begin{align}
\GWlower(X,Y) &= \{(R_\A,R_\B,R_\C): R_\A+R_\C\geq H(X), R_\B+R_\C\notag\\
&\qquad\qquad\;\geq H(Y), R_\A+R_\B+R_\C\geq H(X,Y)\} \label{eq:GWlower}
\end{align}

The Gray-Wyner region was characterized in~\cite{GrayWy74}.
\begin{prop}[\cite{GrayWy74}]\label{prop:GW}
\begin{align*}
&\sRsGW(X,Y) =\\
&\qquad\ihull{\bigcup_{p_{U|X,Y}\in \sP} \{(H(X|U),H(Y|U),I(X,Y;U))\}}
\end{align*}
\end{prop}

The Gray-Wyner system generalizes the setup for Wyner's common
information~\cite{Wyner75} which is defined as the smallest $R_\C$ such
that the outputs of the encoder taken together is an asymptotically
efficient representation of $(X,Y)$, i.e., when $R_\A+R_\B+R_\C=H(X,Y)$.
Using the above proposition we have
\begin{prop}
\label{prop:Wyner}
\begin{align*}
\Com\Wyner(X,Y) &= \inf \{R_\C: (R_\A,R_\B,R_\C)\in \sRsGW(X,Y),\\
       &\qquad\qquad\qquad R_\A+R_\B+R_\C=H(X,Y)\}\\
       &= \inf_{p_{U|X,Y}\in\sP:X-U-Y} I(X,Y;U)
\end{align*}
\end{prop}

\subsection{Known connections}\label{sec:knownfacts}

The following connections between the two systems are known:
\begin{itemize}
\item G\'acs-K\"orner common information can be obtained from
the Gray-Wyner region~\cite[Problem 4.28, pg. 404]{CsiszarKo81}.
\setcounter{MYtempeqncounter1}{\value{equation}}
\begin{align}
\Com \GKW(X,Y) &= \sup \{R_\C: R_\A+R_\C=H(X),R_\B+R_\C\notag\\
   &\qquad\;\;
  =H(Y), (R_\A,R_\B,R_\C)\in \sRsGW\}
\end{align}
Alternatively~\cite{KamathAn10},
\begin{align}
\Com \GKW(X,Y) &= \sup \{R: R \leq I(X;Y),\notag\\ 
   &\qquad\qquad\;\;
   \{R_\C=R\} \cap \GWlower \subseteq \sRsGW\} \label{eq:AltCGKW}
\end{align}

\item Wyner's common information can be obtained from the
G\'acs-K\"orner system~\cite[Corollary 2.3]{PrabhakaranPr10}.
\setcounter{MYtempeqncounter2}{\value{equation}}
\begin{align}
\Com \Wyner(X,Y) = I(X;Y) + \inf_{(R_1,R_2,0)\in\sRsGKW} R_1 + R_2.
\end{align}
\end{itemize}

\section{Relationship between Assisted Common Information and Gray-Wyner Systems}

\begin{thm} \label{thm:affine}
Let $\sRsGW'(X,Y)$ be the image of $\sRsGW(X,Y)$ under the affine map
$f_{X,Y}$ defined below.
\begin{align*}
f_{X,Y}\left(\left[\begin{array}{c}R_\A\\R_\B\\R_\C\end{array}\right]\right)
\defineqq
 \left[\begin{array}{c}R_\A+R_\C-H(X)\\
  R_\B+R_\C-H(Y)\\R_\A+R_\B+R_\C-H(X,Y)\end{array} \right].
\end{align*}
Then
\begin{align*}
\sRsGKW(X,Y) = \ihull{\sRsGW'(X,Y)}.
\end{align*}
\end{thm}

Thus, the assisted residual information region $\sRsGKW(X,Y)$ is the
increasing hull of the Gray-Wyner region $\sRsGW(X,Y)$ under an affine map
$f_{X,Y}$. The map, in fact, computes the gap of $\sRsGW(X,Y)$ to the
simple lower bound $\GWlower(X,Y)$ of \eqref{eq:GWlower} under a coordinate
transformation. The first coordinate of $\sRsGW'$ is indeed the gap between
the (sum) rate at which the first decoder in the Gray-Wyner system receives
data and the minimum possible rate at which it may receive data so that it
can losslessly reproduce $X^n$. The second coordinate has a similar
interpretation with respect to the second decoder. The third coordinate is
the gap between the rate at which the encoder sends data and the minimum
possible rate at which it may transmit to allow both decoders to losslessly
reproduce their respective sources.

It must, however, be noted that the Gray-Wyner region itself does not
possess the monotonicity property of $\sRsGKW$ which leads to
\theoremref{secure-realization-rate} and is therefore less-suited for the
cryptographic application which motivated \cite{PrabhakaranPr10}.  

The two points we noted in \sectionref{knownfacts} fall out of
\theoremref{affine}.

\begin{corol}\label{cor:CGKWfromThm}
\setcounter{MYtempeqncounter}{\value{equation}}
\setcounter{equation}{\value{MYtempeqncounter1}}
\begin{align}
\Com \GKW(X,Y) &= \sup \{R_\C: R_\A+R_\C=H(X),R_\B+R_\C\notag\\
   &\quad\;\;\;
  =H(Y), (R_\A,R_\B,R_\C)\in \sRsGW(X,Y)\}\\
  &= \sup \{R: R \leq I(X;Y),\notag\\ 
   &\qquad\quad
   \{R_\C=R\} \cap \GWlower(X,Y) \subseteq \sRsGW(X,Y)\}
\end{align}
\setcounter{equation}{\value{MYtempeqncounter}}
\end{corol}
\begin{corol}\label{cor:CGWfromThm}
\setcounter{MYtempeqncounter}{\value{equation}}
\setcounter{equation}{\value{MYtempeqncounter2}}
\begin{align}
\Com \Wyner(X,Y) = I(X;Y) + \inf_{(R_1,R_2,0)\in\sRsGKW(X,Y)} R_1 + R_2.
\end{align}
\setcounter{equation}{\value{MYtempeqncounter}}
\end{corol}

Analogous to the definition of $R_\RDO$, we define the axes intercepts on
the other two axes.
\begin{align*}
R_{1-0} &\defineqq \inf\{R_1:(R_1,0,0)\in\sRsGKW\}\\
R_{2-0} &\defineqq \inf\{R_2:(0,R_2,0)\in\sRsGKW\}
\end{align*}
$R_{1-0}$ (resp., $R_{2-0}$) is the rate at which the genie must
communicate when it has a link to only the user who receives $X$ (resp.
$Y$) source so that the users can produce a common random variable
conditioned on which the sources are independent\footnote{Though the
definition allows for zero-rate communication to the other user and
a zero-rate (but non-zero) residual conditional mutual information, it can
be shown from the expression for these rates in
\eqref{eq:corner1}-\eqref{eq:corner2} that there is a scheme which achieves
exact conditional independence and requires no communication to the other
user.}. Using \propositionref{GKW} we can show that
\begin{align}
R_{1-0} &= \inf_{p_{U|X,Y}\in\sP:I(X;U|Y)=I(X;Y|U)=0} I(Y;U|X), 
\label{eq:corner1}\\
R_{2-0} &= \inf_{p_{U|X,Y}\in\sP:I(Y;U|X)=I(X;Y|U)=0} I(X;U|Y).
\label{eq:corner2}
\end{align}
These quantities were identified in~\cite{WolfWu08} and shown to posses a
monotonic property in the context of secure two-party sampling (a result
which \cite{PrabhakaranPr10} generalized).

As we will show below, this pair of quantities is closely related to a pair
which has been identified elsewhere in the context of lossless coding with
side-information~\cite{MarcoEf09} and the Gray-Wyner
system~\cite{KamathAn10}.
Let (following the notation of~\cite{KamathAn10}) 
\begin{align*}
&G(Y\rightarrow X)\\
&\;
=\inf\{R_\C:(H(X|Y),H(Y)-R_\C,R_\C) \in\sRsGW(X,Y)\},\\
&G(X\rightarrow Y)\\
&\;
=\inf\{R_\C:(H(X)-R_\C,H(Y|X),R_\C) \in \sRsGW(X,Y)\}.
\end{align*}
It has been shown~\cite{MarcoEf09,KamathAn10} that $G(Y\rightarrow X)$ is
the smallest rate at which side-information $Y$ may be coded and sent
to a decoder which is interested in recovering $X$ with asymptotically
vanishing probability of error if the decoder receives $X$ coded and sent
at a rate of only $H(X|Y)$ (which is the minimum possible rate which will
allow such recovery). Further, \cite{KamathAn10} arrives at the maximum of
$G(Y\rightarrow X)$ and $G(X\rightarrow Y)$ as a dual to the alternative
definition of $\Com \GKW$ in \eqref{eq:AltCGKW} from the Gray-Wyner system.

We have the following relationship between the two pairs of quantities.
\begin{corol}\label{cor:cornerconnection}
\begin{align}
G(Y\rightarrow X) &= I(X;Y) + R_{1-0},\label{eq:cornerresult1}\\
G(X\rightarrow Y) &= I(X;Y) + R_{2-0}.\label{eq:cornerresult2}
\end{align}
Further,
\begin{align}
&\inf\{R:R\geq I(X;Y), (R_\C=R)\cap\GWlower(X,Y) \subseteq \sRsGW(X,Y)\}\notag\\
&\qquad=\max(G(Y\rightarrow X),G(X\rightarrow Y))\label{eq:kamathdual}\\
&\qquad=I(X;Y)+\max(R_{1-0},R_{2-0})\label{eq:kamathdualresult}.
\end{align}
\end{corol}

\section{Cryptographic application}

The cryptographic problem we consider is of 2-party {\em secure sampling}:
Alice and Bob should sample correlated random variables $(U,V)$
(Alice getting $U$ and Bob getting $V$), such that Alice's view during the
sampling protocol reveals nothing more to her about Bob's outcome $V$
than what her own outcome $U$ reveals to her, and similarly Bob's view
reveals nothing more about Alice's outcome than is revealed by his own
outcome. This is an important special case of {\em secure multi-party computation},
a central problem in modern cryptography. 

However, it is well-known (see for instance
\cite{Wullschleger08thesis} and references therein) that very few distributions can be sampled from in this way,
unless the computation is aided by a {\em set up} --- some correlated random
variables that are given to the parties at the beginning of the protocol.
The set up itself will be from some distribution $(X,Y)$ (Alice gets $X$ and
Bob gets $Y$) which is different from the desired distribution $(U,V)$.
The fundamental question then is, which set ups $(X,Y)$ can be used to securely
sample which distributions $(U,V)$, and {\em how efficiently}. 

We 
restrict ourselves to the setting of {\em
honest-but-curious} players. In this case, the requirements on a protocol  $\Pi$
for securely sampling $(U,V)$ given a set up $(X,Y)$ can be stated as follows, in terms of the outputs and
the views of the parties from the protocol:%
\footnote{Here we state the conditions for ``perfect security,'' but 
our definitions and results generalize to the setting of ``statistical
security,'' where a small statistical error is allowed.}
%
\begin{equation*}
(\Pialiceout(X,Y),\Pibobout(X,Y))  = (U,V)
\end{equation*}
\begin{equation*}
\Pialice(X,Y) \leftrightarrow \Pialiceout(X,Y) \leftrightarrow \Pibobout(X,Y) 
\end{equation*}
\begin{equation*}
\Pialiceout(X,Y) \leftrightarrow \Pibobout(X,Y) \leftrightarrow \Pibob(X,Y) 
\end{equation*}
These three conditions correspond to correctness, security against a curious
Alice and security against a curious Bob, respectively.

In \cite{PrabhakaranPr10}, we showed that the region $\sRsGKW$ can be used as a
measure of cryptographic complexity of correlated random variables 
(a smaller region $\sRsGKW$ corresponding to a higher complexity),
in that the rate at which a pair $(U,V)$ can be securely sampled
given a set up $(X,Y)$ can be upperbounded by the ratio of their
complexity measures.
More formally, there we presented the following result. (For completeness, 
a proof is provided in the appendix.) 
\begin{thm}[\cite{PrabhakaranPr10}]
\label{thm:secure-realization-rate}
If $n_1$ independent copies of a pair of correlated random variables $(U,V)$
can be securely realized from $n_2$ independent copies of a pair of
correlated random variables $(X,Y)$, then $n_1 \sRsGKW(X,Y) \subseteq 
n_2 \sRsGKW(U,V)$ (where multiplication by $n$
refers to $n$-times repeated Minkowski sum). 
\end{thm}


In \cite{PrabhakaranPr10} we gave an instance of pairs $(U,V)$ and $(X,Y)$ such that
the upperbound on the rate at which instances of $(U,V)$ can be securely
sampled from instances of $(X,Y)$ that is implied by the above result
strictly improved on the upperbounds that could be derived from previous
results. These pairs were contrived to highlight the shortcomings of prior
work. Here we give yet another example where the upperbound from our result
strictly improves on prior work, but is further interesting for two reasons:
firstly, the new example is based on natural correlated random variables that
are widely studied (namely, variants of oblivious transfer), and secondly,
the new upperbound we can prove actually matches an easy lowerbound and is
therefore tight.

\subsection{A New Example} \label{sec:example}

We now discuss the new example where our upperbound is not only strictly better
than the previously best available upperbound, but is also tight. 

\begin{eg} Let $S_{A,1},S_{A,2},S_{B,1},S_{B,2} \in \{0,1\}^L$ and $C_A,C_B
\in \{1,2\}$ be six independent random variables all of which are uniformly
distributed over their alphabets.  Consider a pair of random variables
$X,Y$ defined as $X=(C_A,S_{A,1},S_{A,2},S_{B,C_A})$ and
$Y=(C_B,S_{B,1},S_{B,2},S_{A,C_B})$.  Notice that these are in fact a pair
of independent string-oblivious transfers (string-OT's)
of string length $L$ in opposite directions. Let
$U,V$ be a pair of random variables whose joint distribution is the same as
that of $X,Y$, but with $L=1$. In other words, $U,V$ are a pair of
independent bit-OT's in opposite directions. The goal is to characterize
the efficiency with which we may securely generate independent instances of
$U,V$ from independent instances of $X,Y$ for $L>1$. Here efficiency is the
supremum of $n_2/n_1$ over secure sampling schemes which produce $n_2$
independent copies of $(U,V)$ from $n_1$ independent copies of $(X,Y)$.

It is easy to see that $\KK XY$ intersects the co-ordinate axes at
$(1+L,0,0)$, $(0,1+L,0)$, and $(0,0,2L)$. From, these we can immediately
obtain the upperbound of~\cite{WolfWu08} on the efficiency, namely
$(1+L)/2$. Notice that this is dependent on $L$ and would suggest that
(several) long string-OT pairs can be turned into several (more) bit-OT
pairs. However, as we show below, the efficiency of conversion is just 1,
i.e., the best one can do is to turn each pair of string-OT's into a pair
of bit-OT's.

We will show that $\inf\{R_1+R_2:(R_1,R_2,0) \in \KK UV\}= 2$. But,
$(1,1,0)\in \KK XY$. This can be seen by setting
$Q=(C_A,C_B,S_{A,C_B},S_{B,C_A})$ for which $(R_1,R_2,R_\RD)=(1,1,0)$.
Thus, $\inf\{R_1+R_2:(R_1,R_2,0)\in \KK XY\}\leq 2$. Hence, from
\theoremref{secure-realization-rate}, we may conclude that the efficiency
of conversion we are after is 1.

It only remains to characterize $\inf\{R_1+R_2: (R_1,R_2,0)\in\KK UV\}$.
The following lemma, which is proved in the appendix, provides the
required characterization.
\begin{lem}\label{lem:partofexampleinappendix}
\begin{align*}
\inf\{R_1+R_2:(R_1,R_2,0)\in \KK UV\}=2.
\end{align*}
\end{lem}

\end{eg}

\section*{Acknowledgements}

The authors would like to gratefully acknowledge discussions with Venkat
Anantharam, P\'eter G\'{a}cs, and Young-Han Kim. The example in
\sectionref{example} is based on a suggestion by J\"{u}rg Wullschleger.


\appendix

\begin{figure*}[!t]
\normalsize
\setcounter{MYtempeqncounter}{\value{equation}}
\setcounter{equation}{19}
\begin{align}
I(Y;U|X) &= I(X,Y;U) - I(X;U) = H(X|U) + I(X,Y;U) -
H(X),\label{eq:MIeqs1}\\
I(X;U|Y) &= I(X,Y;U) - I(Y;U) = H(Y|U) + I(X,Y;U) - H(Y),\text{ and}\\
I(X;Y|U) &= H(X|U) + H(Y|U) - H(X,Y|U) 
          = H(X|U) + H(Y|U) + I(X,Y;U) - H(X,Y).\label{eq:MIeqs3}
\end{align}
\setcounter{equation}{\value{MYtempeqncounter}}
\hrulefill
\vspace*{4pt}
\end{figure*}
\begin{IEEEproof}[Proof of \propositionref{GacsKo}]

\GKW common information $\Com \GKW$ is defined as the supremum of the set
of $R$ such that for every $\epsilon>0$ there are maps $g_1:\sX^n
\rightarrow \bbZ$, and $g_2:\sY^n \rightarrow \bbZ$ for a sufficiently
large $n$ which satisfy 
\begin{align} 
&\Pr\left( g_1(X^n) \neq g_2(Y^n) \right) \leq \epsilon,\\ 
&\frac{1}{n} H(g_1(X^n))) \geq R - \epsilon.
\end{align}

An alternative defintion which allows for a genie with zero-rate links to
the users is given below. It is easy to see that this can only lead to a
larger value. But as we will show, the definitions are in fact equivalent. 

Let $\Com \GKW '$ be the supremum of the set of $R$ such that for
every $\epsilon>0$ there are maps $f_k:\sX^n\times\sY^n \rightarrow
\{1,\ldots, 2^{n\epsilon}\}$, $(k=1,2)$, $g_1:\sX^n\times\{1,\ldots,
2^{n\epsilon}\} \rightarrow \bbZ$, and $g_2:\sY^n\times\{1,\ldots,
2^{n\epsilon}\} \rightarrow \bbZ$ for a sufficiently large $n$ which
satisfy \eqref{eq:prob-of-error} and 
\begin{align*}
\frac{1}{n}H(g_1(X^n,f_1(X^n,Y^n))) \geq R -\epsilon.
\end{align*}
Clearly, $\Com \GKW' \geq \Com \GKW$.
We first show 
\begin{align}
 \Com \GKW' = I(X;Y) - R_\RDO.\label{eq:CGKWprimefromRDO}
\end{align}
Let $U=g_1(X^n(f_1(X^n,Y^n)))$. Then
\begin{align*}
I(X^n&;Y^n|U)+H(U)\\ &= I(X^n;Y^n|U) + I(X^n,Y^n;U)\\
&= H(X^n,Y^n) - H(X^n|Y^n,U) - H(Y^n|X^n,U)\\
&= I(X^n;Y^n) + I(X^n;U|Y^n) + I(Y^n;U|X^n)\\
&\geq nI(X;Y).
\end{align*}
Therefore, if the maps satisfy \eqref{eq:RDrate}, then
\begin{align*}
H(U) &\geq  nI(X;Y) - I(X^n;Y^n|U)\\
&\geq nI(X;Y) - n(R_\RD + \epsilon)\\
&= n(I(X;Y) - R_\RD - \epsilon)
\end{align*}
which implies \eqref{eq:CGKWprimefromRDO}.

With $\Com \GKW$ replaced by $\Com \GKW'$, we can prove
\eqref{eq:RDO}-\eqref{eq:CGKW} as follows: \eqref{eq:RDO} follows from
\propositionref{GKW}; \eqref{eq:RDO} and \eqref{eq:CGKWfromRDO} imply
\eqref{eq:CGKW}. See \cite[section~II.B]{PrabhakaranPr10} for a proof from
\eqref{eq:CGKW} of the explicit characterization stated at the end of the
proposition. Since this explicit form can be achieved without any
communication from the genie, it follows that $\Com \GKW' = \Com \GKW$.

\end{IEEEproof}

\begin{IEEEproof}[Proof of \theoremref{affine}]

It is easy to prove the above theorem from the single-letter expressions
for the regions in propositions~\ref{prop:GKW} and~\ref{prop:GW} by making
use of the mutual information equalities
\eqref{eq:MIeqs1}-\eqref{eq:MIeqs3} at the top of the page.

\end{IEEEproof}

\begin{IEEEproof}[Proof of \corollaryref{CGKWfromThm}]
\begin{align*}
&\sup\{R_\C:R_\A+R_\C=H(X),\\
 &\qquad\qquad 
  R_\B+R_\C=H(Y), (R_\A,R_\B,R_\C)\in\sRsGW\}\\
&\namedeqq{a}\sup\{R:(0,0,I(X;Y)-R)\in\sRsGW'\}\\
&\namedeqq{b}\sup\{R:(0,0,I(X;Y)-R)\in\sRsGKW\},
\end{align*}
where (a) follows from the definition $\sRsGW'=f(\sRsGW)$. The $\leq$
direction of (b) follows directly from \theoremref{affine}. But $<$ cannot
hold since if $(0,0,I(X;Y)-R)\in\sRsGKW$, then there is a $R'\geq R$ such
that $(0,0,I(X;Y)-R')\in\sRsGW'$. Finally, (c) follows from
\propositionref{GacsKo}.

To arrive at the alternative form, we verify the equivalence of the two
forms.
\begin{align*}
&\{R: R \leq I(X;Y),
   \{R_\C=R\} \cap \GWlower \subseteq \sRsGW\}\\
&\,= \{R_\C: R_\A+R_\C=H(X),\\
   &\qquad\qquad\,
  R_\B+R_\C=H(Y), (R_\A,R_\B,R_\C)\in \sRsGW\}.
\end{align*}
$\subseteq$: if $R \leq I(X;Y)$, then $(H(X)-R,H(Y)-R,R)\in
\{R_\C=R\}\cap\GWlower$.\\
$\supseteq$: Let $s=(H(X)-R_\C,H(Y)-R_\C,R_\C)\in\sRsGW$. Then (a) $R_C\leq
I(X;Y)$ since $s\in\GWlower$, and (b) if $s'=(r_\A,r_\B,R_\C)\in\GWlower$,
then since $r_\A \geq H(X)-R_\C$ and $r_\B\geq H(Y)-R_\C$, we have $s'\geq
s$ (component-wise) which implies that $s'\in\sRsGW$ from the
definition of the \GW system.
\end{IEEEproof}
\begin{IEEEproof}[Proof of \corollaryref{CGWfromThm}]
\begin{align*}
\Com\Wyner &= \inf \{R_\C: (R_\A,R_\B,R_\C)\in \sRsGW,\\
       &\qquad\qquad\qquad R_\A+R_\B+R_\C=H(X,Y)\}\\
  &\namedeqq{a} \inf \{R_1+R_2+I(X;Y): (R_1,R_2,0)\in\sRsGW'\}\\
  &\namedeqq{b} \inf \{R_1+R_2+I(X;Y): (R_1,R_2,0)\in\sRsGKW\},
\end{align*}
where (a) follows from the definition $\sRsGW'=f(\sRsGW)$; (b) follows from
\theoremref{affine}: $\geq$ direction follows directly from the theorem.
But $>$ cannot hold, since by the theorem, if $(R_1,R_2,0)\in\sRsGKW$ then
there exists $(R_1',R_2',0)\in\sRsGW'$ such that $R_1'\leq R_1$ and
$R_2'\leq R_2$.

\end{IEEEproof}

\begin{IEEEproof}[Proof of \corollaryref{cornerconnection}]
\begin{align*}
&G(Y\rightarrow X)\\
&\quad=\inf\{R_\C:(H(X|Y),H(Y)-R_\C,R_\C) \in\sRsGW\},\\
&\quad\namedeqq{a}\inf\{R:(R-I(X;Y),0,0)\in\sRsGW'\}\\
&\quad\namedeqq{b}\inf\{R:(R-I(X;Y),0,0)\in\sRsGKW\}\\
&\quad\namedeqq{c}I(X;Y)+R_{1-0},
\end{align*}
where (a) follows from $\sRsGW'=f(\sRsGW)$. (b) is a consequence of
\theoremref{affine}: And (c) follows from the definition of $R_{1-0}$.

Similarly we get \eqref{eq:cornerresult2}. The equality
\eqref{eq:kamathdual} is proved in~\cite{KamathAn10} which along
with \eqref{eq:cornerresult1}-\eqref{eq:cornerresult2} implies
\eqref{eq:kamathdualresult}.

\end{IEEEproof}

\begin{IEEEproof}[Proof of \theoremref{secure-realization-rate}]
The theorem is in fact corollary 3.2 of~\cite{PrabhakaranPr10} which
follows immediately from Theorem~3.1 of \cite{PrabhakaranPr10} and the
following lemma:
\begin{lem}
\label{lem:addition}
Let the pair of random variables $(X_1,Y_1)$ be independent of the pair
$(X_2,Y_2)$. If $X=(X_1,X_2)$ and $Y=(Y_1,Y_2)$, then
\[ \KK XY = \KK {X_1}{Y_1} + \KK {X_2}{Y_2}.\]
\end{lem} 
For completeness, we give a proof of Theorem~3.1 of \cite{PrabhakaranPr10}
below since the proof was not provided there. This also contains a proof of
\lemmaref{addition} (see (d) below). Please refer \cite{PrabhakaranPr10}
for notation and a statement of the theorem being proved below.

We show that under each step of a secure protocol, $\sRsGKW$ can only
grow.\\
{\em (a) Local computation cannot shrink it:} For all random variables with
 $X - Y - Z$, we have $\KK X {YZ} \supseteq \KK X Y$ and $\KK {XY} Z
\supseteq \KK X Y$.

The first set inclusion follows from the fact that for the joint p.m.f.
$p_{X,Y,Z,Q}=p_{X,Y}p_{Z|Y}p_{Q|X,Y}$
\begin{align*}
I(X;Y,Z|Q) &= I(X;Y|Q)\\
I(Q;Y,Z|X) &= I(Q;Y|X)\\
I(X;Q|Y,Z) &= I(X;Q|Y).
\end{align*}
{\em (b) Communication cannot shrink it:} For all random variables $(X,Y)$
and functions $f$ over the support of $X$ (resp, $Y$), we have $\KK X
{(Y,f(X))} \supseteq \KK X Y$ (resp, $\KK {(X,f(Y))} Y \supseteq \KK X Y$).

The first set inclusion follows from the following facts for the joint
p.m.f $p_{X,Y,Z,Q}=p_{X,Y}p_{Z|Y}p_{Q|X,Y}$:
\begin{align*}
I(X;Y,f(X)|Q,f(X))&=I(X;Y|Q,f(X))\\&\leq I(X;Y|Q)\\
I(X;Q,f(X)|Y,f(X))&=I(X;Q|Y,f(X))\\&\leq I(X;Q|Y)\\
I(Y;Q,f(X)|X)&=I(Y;Q|X)
\end{align*} 
{\em (c) Securely derived outputs do not have a smaller region:} For all
random variables $X,U,V,Y$ such that $X - U - V$ and $U-V-Y$, we have $\KK
U V \supseteq \KK {(X,U)} {(Y,V)}$.

This follows from the following facts for (dependent) random variables
$X,Y,U,V,Q$ which satisfy the Markov chains $X - U - V$ and $U - V -Y$:
\begin{align*}
I(X,U;Y,V|Q) &\geq I(U;V|Q),\\
I(X,U;Q|Y,V) &= I(X,U;Q,Y|V) - I(X,U;Y|V)\\
             &\namedeqq{a} (I(U;Q,Y|V) +
I(X;Q,Y|U,V))\\&\qquad\qquad\qquad\qquad\quad- I(X;Y|U,V)\\
             &\geq I(U;Q|V),\\
\intertext{and similarly}
I(Y,V;Q|X,U) &\geq I(V;Q|U),
\end{align*}
where we used $U-V-Y$ to obtain equality (a).\\
{\em (d) Regions of independent pairs add up:}
If $(X,Y)$ is independent of $(U,V)$, we have $\KK
{(X,U)} {(Y,V)} = \KK X Y + \KK U V$. This follows easily from the following
facts:

For the joint p.m.f. $p_{X,Y}p_{U,V}p_{Q_1|X,Y}p_{Q_2|U,V}$, we have 
\begin{align*}
I(X,U;Y,V|Q_1,Q_2)&=I(X;Y|Q_1) + I(U,V|Q_2)\\
I(X,U;Q_1,Q_2|Y,V)&=I(X;Q_1|Y) + I(U;Q_2|V)\\
I(Y,V;Q_1,Q_2|X,U)&=I(Y;Q_1|X) + I(V;Q_2|U)
\end{align*}
And, for the joint p.m.f. $p_{X,Y}p_{U,V}p_{Q|X,Y,U,V}$, we have
\begin{align*}
I(X,U;Y,V|Q)&\geq I(X;Y|Q) + I(U;V|Q)\\
I(X,U;Q|Y,V)&\geq I(X;Q|Y) + I(U;Q|V)\\
I(Y,V;Q|X,U)&\geq I(Y;Q|X) + I(V;Q|U)
\end{align*}

\end{IEEEproof}

\begin{IEEEproof}[Proof of \lemmaref{partofexampleinappendix}]

By Lemma~\ref{lem:addition}, we need only characterize the $\inf\{R_1+R_2:(R_1,R_2,0)\in
\sRsGKW\}$ of one of the pair of independent bit-OT's. Let us denote one
bit-OT by $A,B$: where $A=(S_1,S_2)\in\{0,1\}^2$ uniformly distributed over
its alphabet and $B=(C,S_C)$, where $C\in\{1,2\}$ is independent of $A$ and
uniformly distrbuted over its alphabet. By \propositionref{GKW},
\begin{align*}
&\inf\{R_1+R_2:(R_1,R_2,0)\in\KK AB\}\\
&\quad= \inf_{p_{Q|A,B}\in\sP:I(A;B|Q)=0} I(B;Q|A) + I(A;Q|B)\\ 
&\quad= H(A|B)+H(B|A)\\
&\qquad\quad - \sup_{p_{Q|A,B}\in\sP:I(A;B|Q)=0} H(A|Q,B) + H(B|Q,A).
\end{align*}
We show below that the sup term is 1. Since $H(A|B)+H(B|A)=2$, this will
allow us to conclude that the smallest sum-rate of $\sRsRD(0)$ of $A,B$ is
1. Invoking the lemma above, the corresponding smallest sum-rate
for $U,V$ is then 2 as required.

To show that the sup term is 1, notice that the only valid choices of
$p_{Q|A,B}$ are such that $I(A;B|Q)=0$. This means that the resulting
$p_{A,B|Q}(.,.|q)$ must belong to one of eight possible classes shown in
\figureref{example-cond-pmf} (for any $q$ with non-zero probability
$p_Q(q)$; we may assume that all $q$'s have non-zero probability without
loss of generality). Recall that there is a cardinality bound on $Q$; let
us denote the alphabet of $Q$ by $\{q_1,q_2,\ldots,q_N\}$, where $N$ is the
cardinality bound. 

\begin{figure*}
\centering
\subfloat[]{\scalebox{0.4}{\includegraphics{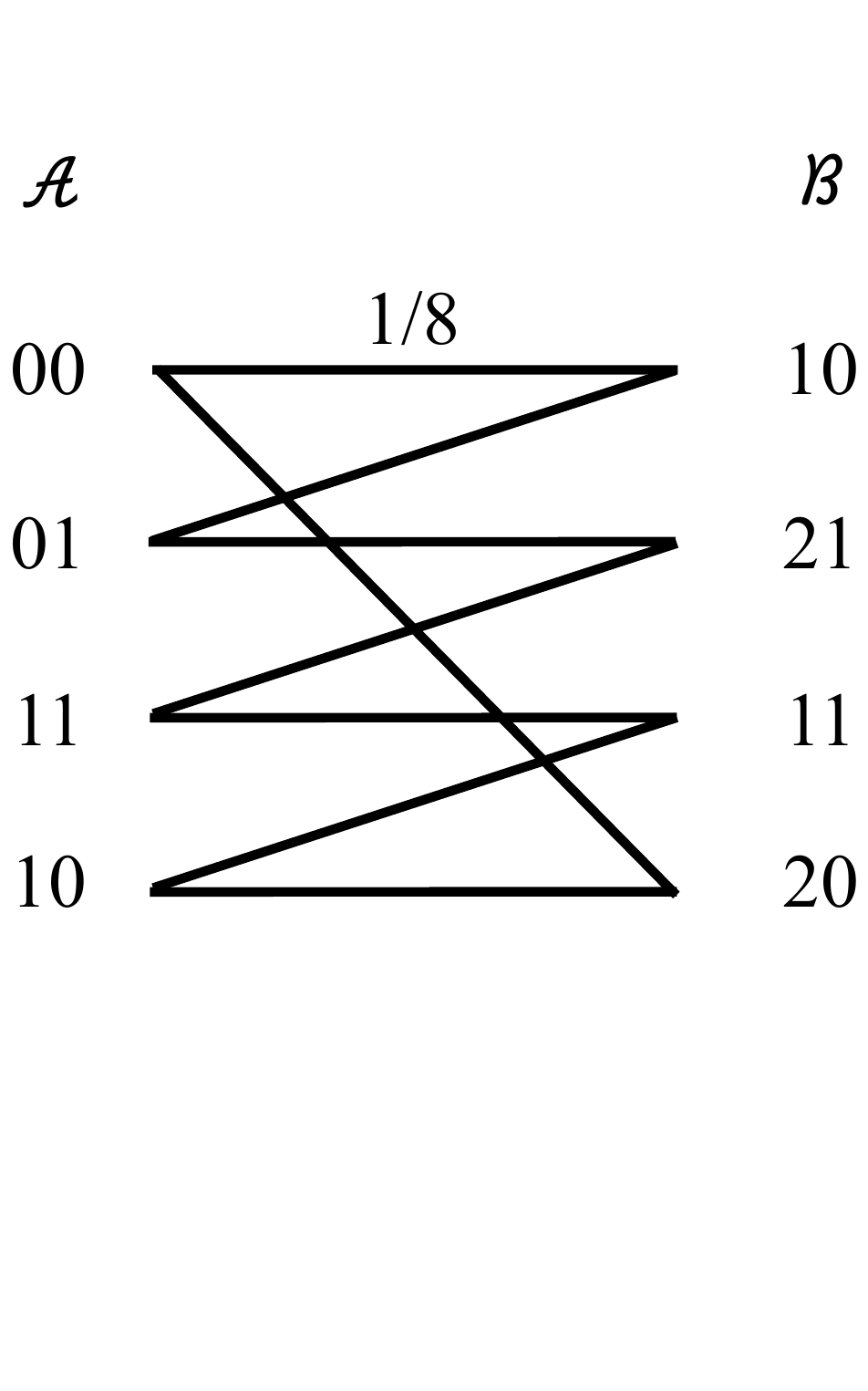}}}\qquad\qquad%
\subfloat[]{\scalebox{0.35}{\includegraphics{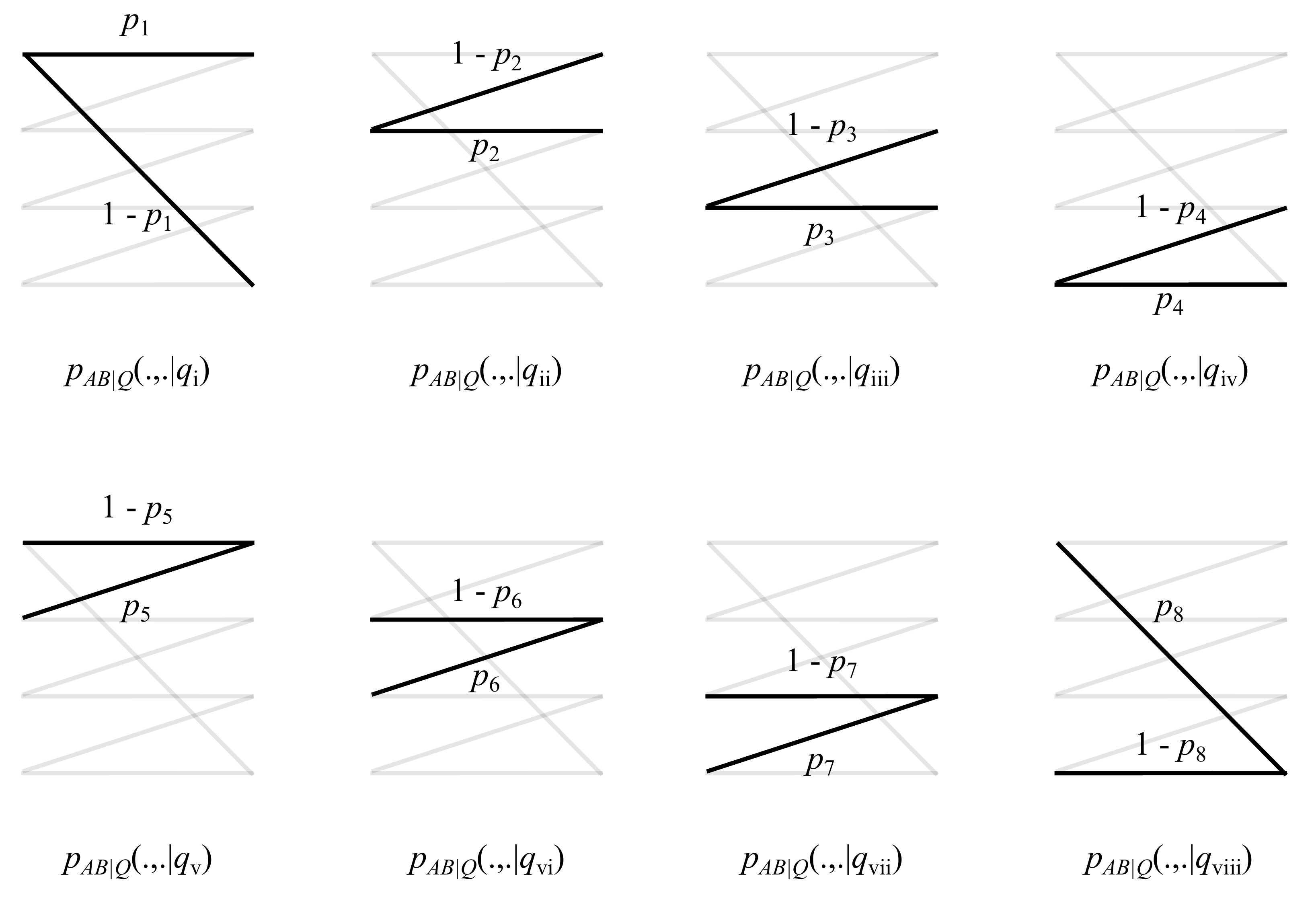}}%
            \label{fig:example-cond-pmf}}
\caption{(a) Joint p.m.f. of $A,B$. Each solid line represents a probablity
mass of 1/8. (b) Eight possible classes that $p_{A,B|Q}(.,.|q)$ may belong
to for a $p_{Q|A,B}$ which satisfies $I(A;B|Q)=0$.}
\label{fig:example}
\end{figure*}

We will first show that there is no loss of generality in assuming that no
more than one of the $q_i$'s is such that its $p_{A,B|Q}(.,.|q_i)$ belongs
to the same class (and hence we may take $N=8$). Suppose, $q_1$ and $q_2$
belong to the same class, say class~1, with parameters $p_1$ and $p_2$
respectively.  Then, if we denote the binary entropy function by $H_2(.)$,
we have
\begin{align*}
&H(A|Q,B) + H(B|Q,A)\\
&= \sum_{k=1}^N p_Q(q_k) \left( H(A|B,Q=q_k) + H(B|A,Q=q_k) \right)\\
&= p_Q(q_1) H_2(p_1) + p_Q(q_2) H_2(p_2)\\
   &\qquad + \sum_{k=3}^N p_Q(q_k) \left(H(A|B,Q=q_k) + H(B|A,Q=q_k)\right)\\
&\leq \left(p_Q(q_1)+p_Q(q_2)\right)
   H_2\left(\frac{p_Q(q_1)p_1 + p_Q(q_2)p_2}{p_Q(q_1)+p_Q(q_2)}\right)\\
  &\qquad + \sum_{k=3}^N p_Q(q_k) \left( H(A|B,Q=q_k)+H(B|A,Q=q_k)\right),
\end{align*}
where the inequality (Jensen's) follows from the concavity of the binary
entropy function. Thus, we can define a $Q'$ of alphabet size $N-1$ where
letters $q_1,q_2$ are replaced by $q_0$ such that
$p_{Q'}(q_0)=p_{Q}(q_1)+p_{Q}(q_2)$, and $p_{A,B|Q'=q_0}$ is in class~1
with parameter $\frac{p_Q(q_1)p_1 + p_Q(q_2)p_2}{p_Q(q_1)+p_Q(q_2)}$, while
maintaining for $i=3,\ldots,N$, $p_{Q'}(q_i)=p_Q(q_i)$ and
$p_{A,B|Q'}(a,b|q_i)=p_{A,B|Q}(a,b|q_i)$. (It is easy to verify (a) that
this gives a valid joint p.m.f. for $p_{A,B,Q'}$, (b) that the induced
$p_{A,B}$ is the same as the original, and (c) that the induced $p_{Q'|A,B}$
satisfies the condition $I(A;B|Q')=0$.) Then, the above inequality states
that
\[ H(A|Q,B) + H(B|Q,A) \leq H(A,Q',B)+H(B|Q',A)\]
proving our claim.

Thus, without loss of generality, we may assume that $N=8$ and
$p_{A,B|Q}(.,.|q_i)$ belongs to class~$i$. Notice that 
\begin{align*}
p_{Q|A,B}(q_1|00,10) + p_{Q|A,B}(q_5|00,10) &= 1,
\\
p_{Q|A,B}(q_2|01,10) + p_{Q|A,B}(q_5|01,10) &= 1,
\\
p_{Q|A,B}(q_2|01,21) + p_{Q|A,B}(q_6|01,21) &= 1,
\\
p_{Q|A,B}(q_3|11,21) + p_{Q|A,B}(q_6|11,21) &= 1,
\\
p_{Q|A,B}(q_3,11,11) + p_{Q|A,B}(q_7|11,11) &=1,
\\
p_{Q|A,B}(q_4|10,11) + p_{Q|A,B}(q_7|10,11) &= 1,
\\
p_{Q|A,B}(q_4|10,20) + p_{Q|A,B}(q_8|10,20) &= 1,
\\
p_{Q|A,B}(q_1|00,20) + p_{Q|A,B}(q_8|00,20) &= 1.
\end{align*}
Let us define 
\begin{align*}
\tp_1 &\defineqq p_{Q|A,B}(q_1|00,10),&&
\tp_5 \defineqq p_{Q|A,B}(q_5|01,10),\\
\tp_2 &\defineqq p_{Q|A,B}(q_2|01,21),&&
\tp_6 \defineqq p_{Q|A,B}(q_6|11,21),\\
\tp_3 &\defineqq p_{Q|A,B}(q_3|11,11),&&
\tp_7 \defineqq p_{Q|A,B}(q_7|10,11),\\
\tp_4 &\defineqq p_{Q|A,B}(q_4|10,20),&&
\tp_8 \defineqq p_{Q|A,B}(q_8|00,20).
\end{align*}

Let us evaluate $H(B|Q,A)$ in terms of the above parameters. Notice that
$H(B|Q=q_i,A)=0$ for $i=5,\ldots,8$. Hence
\begin{align*}
&H(B|Q,A)\\
&=\sum_{\begin{subarray}{c}(q,a)\in\{(1,00),(2,01),\\
                           \qquad\;\;\;\,(3,11),(4,10)\}
        \end{subarray}} p_{Q,A}(q,a)H(B|Q=q,A=a)\\
 &= \frac{\tp_1+(1-\tp_8)}{8}H_2\left(\frac{\tp_1}{\tp_1+(1-\tp_8)}\right)\\
   &\qquad+ 
    \frac{\tp_2+(1-\tp_5)}{8}H_2\left(\frac{\tp_2}{\tp_2+(1-\tp_5)}\right)\\
   &\qquad 
    +\frac{\tp_3+(1-\tp_6)}{8}H_2\left(\frac{\tp_3}{\tp_3+(1-\tp_6)}\right)\\
   &\qquad+ 
    \frac{\tp_4+(1-\tp_7)}{8}H_2\left(\frac{\tp_4}{\tp_4+(1-\tp_7)}\right)\\
 &\leq \frac{4+\sum_{i=1}^4\tp_i - \sum_{j=5}^8 \tp_j}{8},
\end{align*}
where the inequality follows from the fact that binary entropy function is
upperbounded by 1. Similary, we can get
\begin{align*}
H(A|Q,B) \leq \frac{4+\sum_{j=5}^8\tp_j - \sum_{i=1}^4\tp_i}{8}.
\end{align*}
Combining, we obtain the desired
\[ H(B|Q,A) + H(A|Q,B) \leq 1.\]

\end{IEEEproof}

\end{document}